\documentclass[11pt]{article}

\usepackage[a4paper,left=2cm,right=2cm,top=1.5cm,bottom=2cm]{geometry}
\usepackage[utf8]{inputenc}
\usepackage[T1]{fontenc}
\usepackage[english]{babel}
\usepackage{amsthm}
\usepackage{amsmath}
\usepackage{amssymb}
\usepackage{mathrsfs}
\usepackage{setspace}
\usepackage{float}
\usepackage{multirow}
\usepackage{stmaryrd}
\usepackage{color}
\usepackage{wrapfig}
\usepackage{pifont}
\usepackage{array}
\usepackage[colorlinks=true,linkcolor=ocre,citecolor=ocre]{hyperref}
\usepackage{dsfont}
\usepackage{upgreek}
\usepackage{nicefrac}
\usepackage{tikz}{}
\usepackage{gensymb}
\usepackage{FiraSans}

\usepackage{graphicx}
\usepackage{caption}
\renewenvironment{proof}{\noindent{\sffamily{\textbf{Proof :}}}}{\begin{flushright}$\square$\end{flushright}}

\newcommand{\IE}{\mathbb{E}}

\newcommand{\IZ}{\mathbb{Z}}

\newcommand{\IR}{\mathbb{R}}

\newcommand{\IT}{\mathbb{T}}

\newcommand{\drm}{\mathrm d}

\newcommand{\CD}{\mathcal D}

\newcommand{\CC}{\mathcal C}
\newcommand{\CH}{\mathcal H}

\newcommand{\CX}{\mathcal X}

\newcommand{\CB}{\mathcal B}

\renewcommand{\P}{\mathsf{P}}
\newcommand{\PT}{\widetilde{\P}}
\newcommand{\PI}{\mathsf{\Pi}}

\newcommand{\DF}{\mathsf{F}}

\newcommand{\DO}{\mathsf{F}}
\newcommand{\DB}{\mathsf{B}}

\newcommand{\EA}{\mathbf{A}}

\definecolor{ocre}{RGB}{64,123,121}
\definecolor{S}{rgb}{0.0,0.5,0.0}

\newcounter{item}
\numberwithin{item}{section}

\newtheorem{theorem}[item]{\sffamily Theorem}
\newtheorem{definition}[item]{\sffamily Definition}
\newtheorem{proposition}[item]{\sffamily Proposition}
\newtheorem{lemma}[item]{\sffamily Lemma}
\newtheorem{corollary}[item]{\sffamily Corollary}

\newtheorem*{theorem*}{\sffamily Theorem}
\newtheorem*{definition*}{\sffamily Definition}
\newtheorem*{proposition*}{\sffamily Proposition}
\newtheorem*{lemma*}{\sffamily Lemma}
\newtheorem*{corollary*}{\sffamily Corollary}

\usepackage[explicit]{titlesec}
\titleformat{\section}{\centering\Large\bfseries}{\thesection \ --}{0.7em}{\Large\bfseries #1}
\titleformat{\subsection}{\centering\large\bfseries}{\thesubsection \ --}{0.4em}{\large\bfseries #1}
\titleformat{\subsubsection}{\centering\bfseries}{\thesubsubsection \ --}{0.4em}{\bfseries #1}

\let\emph\relax
\DeclareTextFontCommand{\emph}{\bfseries\em}

\providecommand{\MSC}[1]
{
	{\footnotesize	
	\textbf{MSC $\mathbf{2020}$ --} #1}
}

\providecommand{\keywords}[1]
{
	{\footnotesize	
	\textbf{Keywords --} #1}
}

\title{\bfseries 2D random magnetic Laplacian with white noise magnetic field}
\author{Léo MORIN and Antoine MOUZARD}
\date{}

\begin{document}

\maketitle
\abstract{We define the random magnetic Laplacian with spatial white noise as magnetic field on the two-dimensional torus using paracontrolled calculus. It yields a random self-adjoint operator with pure point spectrum and domain a random subspace of nonsmooth functions in $L^2$. We give sharp bounds on the eigenvalues which imply an almost sure Weyl-type law.}

\bigskip

\MSC{35J10; 60H25; 47A52}


\tableofcontents
\vspace{0.5cm}

\section*{Introduction}

The magnetic Laplacian associated to a magnetic potential $A$ in two dimensions
$$
H=(i\partial_1+A_1)^2+(i\partial_2+A_2)^2
$$
is of interest in the description of a number of physical models. For example, it describes the behavior of a particule in a magnetic field $B$ related to $A$ via
$$
B=\nabla\times A=\partial_2A_1-\partial_1A_2.
$$
While the case of constant magnetic field has been largely studied, the analysis of the magnetic Laplacian with nonconstant magnetic field gives rise to a number of interesting questions. Apart from the analogy with the electric Laplacian $-\Delta+V$, the spectral study of the magnetic Laplacian is also motivated by supraconductivity, where $H$ plays a specific role in the third critical field of Ginzburg-Landau theory, see \cite{HelfFour}. This work is dedicated to the study of the magnetic Laplacian with random singular magnetic field given by the space white noise $B=\xi$ on the two-dimensional torus $\IT^2$. It can be constructed as a distribution with independant random Fourier coefficients with centered normal law of unit variance. In two dimensions, the space white noise belongs almost surely to the Sobolev spaces $\CH^{-1-\kappa}$ or the Besov-Hölder spaces $\CC^{-1-\kappa}$ for any $\kappa>0$. Since the associated potential $A$ satisfies the equation
$$
\xi=\partial_2A_1-\partial_1A_2,
$$
each component $A_1,A_2$ is expected to belong to $\CC^{-\kappa}$ for any $\kappa>0$. Thus the potential is not a measurable function and the associated magnetic Laplacian falls out of the range of the classical theory: It is a singular random operator. The study of PDEs or operators where rough stochastic terms give rise to ill-defined operations such as product of distributions has intensified since the $2010$'s with the parallel introduction of regularity structures in \cite{Hai14} by Hairer and the paracontrolled calculus in \cite{GIP} by Gubinelli, Imkeller and Perkowski. A famous example of such singular operator is the Anderson Hamiltonian
$$
-\Delta+\xi
$$
which naturally appears in the parabolic Anderson model describing random walks in random environment, see \cite{Konig} for a survey on the topic. The random operator introduced and studied in this work is the magnetic analog of the Anderson Hamiltonian with a first order perturbation of the Laplacian.

\medskip

The class of singular stochastic PDEs consists of equations with rough stochastic terms which give rise to ill-defined terms, classicaly products of distributions. Examples are given by the KPZ equation
$$
(\partial_t-\partial_x^2)u=\zeta+(\partial_xu)^2
$$
with $\zeta$ a spacetime white noise in dimension $1+1$ or the $\Phi_d^4$ equation
$$
(\partial_t-\Delta)\phi=-\phi^3+\xi
$$
with $\xi$ a space white noise in dimension $d\ge2$. These equations do not make sense a priori since solutions are expected to be too irregular for the terms $(\partial_xu)^2$ or $\phi^3$ to make sense. The recently developped approach consists in the construction of a random subspace of classical function spaces through a stochastic renormalisation procedure from the noise in which one can make sense almost surely of the equation. The different approaches of regularity structures and paracontrolled calculus differ by the tools used to study the singular products. In the first one, distributions are locally described by generalised Taylor expansions while the second one relies on tools from harmonic analysis with a description of distributions by frequency ``schemes''. In both approaches, solutions are described by a richer stochastic object than the noise called the enhanced noise, the renormalisation procedure corresponds to its construction. This idea takes root in the theory of Lyons' rough paths and Gubinelli's controlled paths developped as a pathwise approach to stochastic integration. This has been used to study the Anderson Hamiltonian
$$
-\Delta+\xi
$$
which is singular in dimension $d\ge2$, the first time by Allez and Chouk using paracontrolled calculus on the two-dimensional torus, see \cite{AllezChouk}. See also \cite{GUZ,Labbe,Mouzard} for more general settings including dimension $d\in\{2,3\}$, different boundary conditions or a more general geometrical framework. Following the approach for singular SPDEs, the idea is to construct a random subspace $\CD_\Xi$ of $L^2$ as domain for the operator given an enhancement $\Xi$ of the noise $\xi$ such that $(H,\CD_\Xi)$ is a well-defined unbounded operator in $L^2$. The singularity is dealed with through a renormalisation procedure in the construction of the enhanced noise $\Xi$. In this work, we construct such a domain for the random magnetic Laplacian with white noise magnetic field.

\medskip

Since the Anderson Hamiltonian appears, at least formally, as the continuum limit of discrete Anderson Hamiltonians with independant and identically distributed random potentials with finite variance, the random magnetic Laplacian naturally describes the formal continuous limit of Hamiltonians with magnetic random fields. It is also of interest to the study of different PDEs as for the deterministic magnetic Laplacian. For example, it can be used to solve dispersive PDEs like the Schrödinger equation
$$
\partial_tu=(i\partial_1+A_1)^2u+(i\partial_2+A_2)^2u-|u|^2u
$$
or the wave equation. In particular, see \cite{MZ} where Strichartz inequalities are obtained for the Anderson Hamiltonian as well as the random magnetic Laplacian constructed in the present work.

\bigskip

{\noindent\large\textbf{Main results}}

\medskip

The $2D$ magnetic Laplacian with potential $A:\IT^2\to\IR^2$ is formally given by
$$
H=-\Delta+2iA\cdot\nabla+A\cdot A+i\textup{div}{A}
$$
with associated magnetic field
$$
B=\nabla\times A.
$$
From the physical point of view, one considers a given magnetic field $B:\IT^2\to\IR$ and chooses a potential $A$ satisfying this equation. Different choices of $A$ yield different operators $H_A$, this is the choice of gauge. However, the physical quantities are expected to be independent of this choice and there is a partial invariance by gauge change. Indeed, if $A:\IT^2\to\IR^2$ is a given potential associated to $B$, then the new potential $A+\nabla f$ is still a potential associated to $B$ for functions $f:\IT^2\to\IR$. The two operators are related formally by the relation
$$
H_{A+\nabla f}=e^{if}H_Ae^{-if}
$$
hence the spectral properties of $H_A$ and $H_{A+\nabla f}$ are the same. While on $\IR^2$, every pair $(A,A')$ of smooth potential that satifies
$$
\nabla\times A=\nabla\times A'
$$
are related by
$$
A=A'+\nabla f,
$$
this is no longer the case on $\IT^2$ due to topological holes. One can show that the choice of a potential $A$ can influence the spectrum of $H_A$, see for example the Aharonov-Bohm effect in \cite{Helffer}. In this work, we consider the magnetic field $B=\xi$ the space white noise and we choose the Lorentz gauge defined as follows. Take
$$
A=\nabla^\bot\varphi\quad\text{where}\quad\varphi:=\Delta^{-1}\xi.
$$
Then $A$ satifies indeed
$$
\xi=\nabla\times A
$$
with the addition of $\textup{div}(A)=0$. Our method could also deal with a gauge $A+\nabla f$ with $A$ the Lorentz gauge and $f$ a smooth deterministic function, see Section \ref{Renormalisation} for a discussion about this. It would be interesting to study the influence of a more general gauge but the implict definition
$$
B=\nabla\times A
$$
makes more difficult the study of singularities. This leaves us with 
$$
H=-\Delta+2iA\cdot\nabla+A\cdot A
$$
where $A\in\CC^{\alpha-1}$ almost surely for any $\alpha<1$. The term $A\cdot A$ is singular and one has to give a meaning to it using probabilistic arguments. This is done in Section \ref{Renormalisation} and yields the associated enhanced potential
$$
\EA=(A,A^2)\in\CC^{\alpha-1}(\IT^2,\IR^2)\times\CC^{2\alpha-2}(\IT^2,\IR).
$$
Remark that since $A$ is a distribution of negative Hölder regularity, the singular product $A\cdot A$ is expected to worsen the regularity. Given such an enhanced potential $\EA$, we construct in Section \ref{DefOp} a dense subspace $\CD_\EA\subset L^2$ such that 
$$
u\in\CD_\EA\subset L^2\quad\implies\quad Hu\in L^2.
$$
In Section \ref{SelfAdj}, we show that $(H,\CD_\EA)$ is almost surely a self-adjoint operator with pure point spectrum. We also prove that it is the resolvent-limit of
$$
H_\varepsilon=-\Delta+2iA_\varepsilon\cdot\nabla+A_\varepsilon^2
$$
for any regularisation $\EA_\varepsilon=(A_\varepsilon,A_\varepsilon^2)\in C^\infty(\IT^2,\IR^2)\times C^\infty(\IT^2,\IR)$ such that
$$
\lim_{\varepsilon\to0}\|A-A_\varepsilon\|_{\CC^{\alpha-2}}+\|A^2-A_\varepsilon^2\|_{\CC^{2\alpha-2}}=0.
$$
Finally, we construct in Section \ref{Renormalisation} the enhanced potential $\EA$ associated to the random magnetic field $B=\xi$. In particular, it is described by the limit in probability of
$$
\big(A_\varepsilon,A_\varepsilon^2-c_\varepsilon\big)
$$
as $\varepsilon$ goes to $0$ where $A_\varepsilon$ is a regularisation of $A$ and
$$
c_\varepsilon=\IE\big[A_\varepsilon(0)\cdot A_\varepsilon(0)\big].
$$
In particular, the almost sure singularity of the product $A\cdot A$ implies the need to substract a diverging constant $c_\varepsilon$ as $\varepsilon$ goes to $0$ and a singular random operator has to be interpreted as the description of the limiting behavior of a diverging system. One is interested in the fluctuations of this system in this diverging frame, as the central limit Theorem for a simple random walk. For the case of the Anderson Hamiltonian, see the work \cite{MP1} of Martin and Perkowski for a nice example.

\medskip

Our construction is done in the same spirit as in \cite{Mouzard} for the Anderson Hamiltonian with the additional difficulty of the first order perturbation here. In particular, this work illustrates that the heat paracontrolled calculus deals naturally with such terms and its flexibility to solve singular stochastic PDEs. In particular, this shows that the method could be used to deal with a general class of operators of the form
$$
-\Delta+a_1\cdot\nabla+a_2
$$
with rough stochastic fields $a_1,a_2$ and therefore associated time-dependant PDEs. A very important example of such operators in probability corresponds to the case $a_2=0$ where one recovers the infinitesimal generator of an SDE with drift $a_1$. For example, it makes possible the study of processes as the Brox diffusion in a more general framework, see for example \cite{KP} by Kremp and Perkowski. This is current investigation and will be done in a subsequent paper. Finally since we only define and prove self-adjointness of the operator, no major differences appear in the study of the Anderson Hamiltonian or the random magnetic Laplacian with white noise magnetic field. As one does for smooth potentials, it is natural to pursue the study of both operators to understand their differences. For example, an interesting first question it to investigate the asymptotics of the eigenvalues as the volume of the torus goes to infinity.

\medskip

In both the Anderson and the magnetic case, it would be interesting to study the influence of a scaling $\lambda>0$ in the noise. For instance, the behavior of
$$
-\Delta+\lambda\xi
$$
as $\lambda$ to $0$ should not be a problem however the limit $\lambda$ to $\infty$ is much more delicate and of great interest. It is equivalent to the large-volume limit obtained by taking a torus of large diameter $\sqrt{\lambda}$. It was studied for example in \cite{CZ} by Chouk and Van Zuijlen using the large deviation principle, see also \cite{KPZ} for motivations from Probability. This is also equivalent to the semiclassical limit $h$ goes to $0$ in
$$
-h\Delta+\xi.
$$
For regular potentials $V$, the semiclassical limit is deeply related to the classical dynamics described by the Hamiltonian
$$
(x,v)\in T^*(\IT^2)\mapsto|v|^2+V(x)
$$
with $T^*(\IT^2)$ the cotangent bundle of $\IT^2$. Similarily, would it be possible to find an effective ``classical'' equation describing the semiclassical limit in the case of a singular potential $V=\xi$ or $B=\xi$ ? See the book \cite{Helffer} from Helffer for details on the semiclassical limit and the work \cite{DumazLabbe} by Dumaz and Labbé for a very precise description of the asymptotics in one dimension for the Anderson Hamiltonian.

\bigskip

{\noindent\large\textbf{Organisation of the paper}}

\medskip

In the first section, we construct the domain and prove density in $L^2$. We compare the graph norm and the natural norms of the domain which gives closedness of the operator. We also give an explicit form comparison between the random magnetic Laplacian $H$ and the Laplacian $\Delta$. In the second section, we show that the operator is symmetric as a weak limit of the regularised operator. The form comparison of $H$ and $\Delta$ with the Babuška-Lax-Milgram Theorem gives self-adjointness. Finally, we show that $H$ is the resolvent-limit of the regularised operator $H_\varepsilon$ in Proposition \ref{ResolventConvergence} and compare the spectrum of $H$ and $\Delta$. In particular, this implies an almost sure Weyl-type law for the random magnetic Laplacian. The third Section deals with the construction of the enhanced potential $\EA$ built from the noise $\xi$ through a renormalisation procedure. We gather in Appendix \ref{AppendixPC} all the results we need for the paracontrolled calculus and refer to \cite{Mouzard} for the details. In particular, $\P$ and $\PI$ respectively denotes the paraproduct and the resonent term such that
$$
fg=\P_fg+\P_gf+\PI(f,g)
$$
and $\PT$ is a paraproduct intertwinned with $\P$ via the relation
$$
\Delta\circ\PT=\P\circ \Delta
$$
and satisfies the same continuity properties as $\P$. We also denote
$$
\Delta^{-1}:=-\int_0^1e^{t\Delta}\drm t
$$
an inverse for the Laplacian up to the regularising operator $e^{\Delta}$.

\bigskip

\section{Definition of the operator}\label{DefOp}

In this section, we first construct the domain and show that its natural norms are equivalent to the graph norm of $H$. In particular, this guarantees the closedness of the operator. Finally, we compare the respective forms associated to $H$ and $L$.

\bigskip

\subsection{Construction of the domain}

Fix $\alpha\in(\frac{2}{3},1)$ and let $\EA$ be an enhanced magnetic potential 
$$
\EA=(A,A^2)\in\CX^\alpha:=\CC^{\alpha-1}(\IT^2,\IR^2)\times\CC^{2\alpha-2}(\IT^2,\IR)
$$
with its natural norm
$$
\|\EA\|_{\CX^\alpha}:=\|A\|_{\CC^{\alpha-1}}+\|A^2\|_{\CC^{2\alpha-2}}.
$$
For $A\in L^\infty$, the term $A^2$ can be interpreted as $A\cdot A$ while it is not defined if $A$ is only a distribution. It is enhanced in the sense that one does not have a natural interpretation of $A\cdot A$ for singular potential, this is specified by the additional data $A^2$. Section \ref{Renormalisation} is devoted to the particular case of magnetic white noise where $A^2$ is constructed through a probabilistic renormalisation procedure. Thus we refer as noise-dependent a quantity that depends on this enhanced potential $\EA$.

\bigskip

For any regular function $u\in C^\infty(\IT^2)$, we have
$$
2iA\cdot\nabla u+A^2u=\P_{\nabla u}2iA+\P_u A^2+(\sharp)
$$
where $(\sharp)\in C^\infty(\IT^2)$ with $\P_u2iA\in\CH^{\alpha-1}$ and $\P_{\nabla u}A^2\in\CH^{2\alpha-2}$. Assuming $Hu\in L^2$ yields
$$
-\Delta u=Hu-2iA\cdot\nabla u+A^2u\in\CH^{2\alpha-2}
$$
since $2\alpha-2<\alpha-1$ hence $u$ is expected to belong to $\CH^{2\alpha}$ by elliptic regularity theory. For $u\in\CH^{2\alpha}$, we have
\begin{align*}
2iA\cdot\nabla u+A^2u&=2i\P_{\nabla u}A+2i\P_A\nabla u+2i\PI(\nabla u,A)+\P_uA^2+\P_{A^2}u+\PI(u,A^2)\\
&=(\alpha-1)+(3\alpha-2)+(2\alpha-2)+(4\alpha-2)\\
&=\P_uA^2+\P_{\nabla u}2iA+(3\alpha-2)
\end{align*}
where $(\beta)$ denotes a term of formal regularity $\CH^\beta$ for any $\beta\in\IR$. Following the paracontrolled calculus approach, we want to consider a paracontrolled function of the form 
$$
u=\PT_uX_1+\PT_{\nabla u}X_2+u^\sharp
$$ 
with $u^\sharp$ a smoother remainder such that $Hu\in L^2$. Thus we take
$$
X_1:=\Delta^{-1}(A^2)\quad\text{and}\quad X_2:=\Delta^{-1}(2iA)
$$
and define the domain of $H$ as follows.

\medskip

\begin{definition*}
We define the set $\CD_\EA$ of functions paracontrolled by $\EA$ as
$$
\CD_\EA:=\{u\in L^2;\ u-\PT_uX_1-\PT_{\nabla u}X_2\in\CH^2\}.
$$
\end{definition*}

\medskip

The domain is defined as
$$
\CD_\EA=\Phi^{-1}(\CH^2)
$$
with
$$
\Phi(u):=u-\PT_uX_1-\PT_{\nabla u}X_2.
$$
However the domain could be anything from trivial to dense in $L^2$ a priori. For $s\in(0,1)$, we introduce the map $\Phi^s$ as
$$
\Phi^s:\left|\begin{array}{ccc}
\CD_\EA&\to&\CH^2\\
u&\mapsto&u-\PT_u^sX_1-\PT_{\nabla u}^sX_2
\end{array}\right.
$$
with $\PT^s$ the paraproduct truncated at scale $s$; see Appendix \ref{AppendixPC} for the definition. In particular, the map
$$
\Phi^s:\CH^\beta\to\CH^\beta
$$
is a perturbation of the identity for any $\beta\in[0,2\alpha)$ invertible for $s$ small enough, we denote its inverse $\Gamma$. Since $\big(\PT_v-\PT_v^s\big)X$ is a smooth function, the domain is also given by
$$
\CD_\EA=(\Phi^s)^{-1}(\CH^2)=\Gamma(\CH^2).
$$
The reader should keep in mind that $\Gamma$ implicitely depends on $s$, we do not keep it in the notation to lighten this work. This parametrisation of the domain will be crucial to prove that the domain is dense in $L^2$ and to study $H$. In particular, sharp bounds on the eigenvalues of $H$ are needed to get a Weyl-type law for $H$. To do so, we need to keep a careful track of the different constants. The reader interested only in the construction of the operator and its self-adjointess can skip these computations. Obtaining sharp bounds requires explicit constants with respect to the parameter $s$ and the regularity exponent in the paracontrolled calculus. We recall the needed bounds in Appendix \ref{AppendixPC}, see \cite{Mouzard} for details and proofs. For $\beta\in[0,2\alpha)$, let
$$
s_\beta(\EA):=\left(\frac{\beta^*}{m\|\EA\|_{\CX^\alpha}}\right)^{\frac{4}{2\alpha-\beta}}
$$
where $\beta^*=1-\beta$ if $\beta\in[0,1)$ and $\beta^*=2\alpha-\beta$ if $\beta\in[1,2\alpha)$ and $m>0$ is a universal constant, see Appendix \ref{AppendixPC}. The following Proposition gives regularity estimates for $\Phi^s$ and $\Gamma$. 

\medskip

\begin{proposition}\label{GammaBound}
Let $\beta\in[0,2\alpha)$ and $s\in(0,1)$. We have 
$$
\|\Phi^s(u)-u\|_{\CH^\beta}\le m\frac{s^{\frac{2\alpha-\beta}{4}}}{\beta^*}\|\EA\|_{\CX^\alpha}\|u\|_{\CH^\beta}.
$$
In particular, $s<s_\beta(\EA)$ implies that the map $\Phi^s$ is invertible and its inverse $\Gamma$ verifies the bound
$$
\|\Gamma u^\sharp\|_{\CH^\beta}\le\frac{1}{1-m\frac{s^{\frac{2\alpha-\beta}{4}}}{\beta^*}\|\EA\|_{\CX^\alpha}}\|u^\sharp\|_{\CH^\beta}.
$$
\end{proposition}

\medskip

\begin{proof}
If $\beta<1$, the bounds on $\Phi^s$ follow directly from
\begin{align*}
\|\PT_u^sX_1+\PT_{\nabla u}^sX_2\|_{\CH^\beta}&\le m\frac{s^{\frac{2\alpha-\beta}{4}}}{1-\beta}\|u\|_{L^2}\|X_1\|_{\CC^{2\alpha}}+m\frac{s^{\frac{\alpha+1-\beta}{4}}}{1-\beta}\|\nabla u\|_{\CH^{\beta-1}}\|X_2\|_{\CH^{\alpha+1}}\\
&\le m\frac{s^{\frac{2\alpha-\beta}{4}}}{1-\beta}\|u\|_{\CH^\beta}\big(\|X_1\|_{\CC^{2\alpha}}+\|X_2\|_{\CC^{\alpha+1}}\big)\\
&\le m\frac{s^{\frac{2\alpha-\beta}{4}}}{1-\beta}\|u\|_{\CH^\beta}\|\EA\|_{\CX^\alpha}.
\end{align*}
For $\beta\in[1,2\alpha)$, we have
\begin{align*}
\|\PT_u^sX_1+\PT_{\nabla u}^sX_2\|_{\CH^\beta}&\le m\frac{s^{\frac{2\alpha-\beta}{4}}}{2\alpha-\beta}\|u\|_{L^2}\|X_1\|_{\CC^\alpha}+m\frac{s^{\frac{\alpha+1-\beta}{2}}}{\alpha+1-\beta}\|\nabla u\|_{L^2}\|X_2\|_{\CH^{\alpha+1}}\\
&\le m\frac{s^{\frac{2\alpha-\beta}{4}}}{2\alpha-\beta}\|u\|_{\CH^1}\big(\|X_1\|_{\CC^{2\alpha}}+\|X_2\|_{\CC^{\alpha+1}}\big)\\
&\le m\frac{s^{\frac{2\alpha-\beta}{4}}}{2\alpha-\beta}\|u\|_{\CH^1}\|\EA\|_{\CX^\alpha}.
\end{align*}
The result for $\Gamma$ follows since $2\alpha-\beta>0$.
\end{proof}

\medskip

We also consider the associated maps $\Phi_\varepsilon^s$ and $\Gamma_\varepsilon$ for a regularisation $\EA_\varepsilon$ of the enhanced potential. It is defined as
$$
\Phi_\varepsilon^s(u):=u-\PT_uX_1^{(\varepsilon)}-\PT_{\nabla u}X_2^{(\varepsilon)}
$$
and
$$
\Gamma_\varepsilon u^\sharp=\PT_{\Gamma_\varepsilon u^\sharp}X_1^{(\varepsilon)}+\PT_{\nabla\Gamma_\varepsilon u^\sharp}X_2^{(\varepsilon)}+u^\sharp
$$
where
$$
X_1^{(\varepsilon)}:=\Delta^{-1}(A_\varepsilon^2)\quad\text{and}\quad\Delta X_2^{(\varepsilon)}:=2iA_\varepsilon.
$$
These satisfy the same continuity properties as $\Phi^s$ and $\Gamma$ with bounds uniform with respect to $\varepsilon$. Moreover, we have the following approximation Lemma holds.

\medskip

\begin{lemma}\label{LemmaGammaConvergence}
Let $\beta\in[0,2\alpha)$ and $s\in(0,1)$. If $s\le s_\beta(\EA)$, we have
$$
\|\textup{Id}-\Gamma\Gamma_{\varepsilon}^{-1}\|_{L^2\to\CH^\beta}\lesssim_{\EA,s}\|\EA-\EA_\varepsilon\|_{\CX^\alpha}.
$$
In particular, this implies the convergence of $\Gamma_\varepsilon$ to $\Gamma$ with the bound
$$
\|\Gamma-\Gamma_\varepsilon\|_{\CH^\beta\to\CH^\beta}\lesssim_{\EA,s}\|\EA-\EA_\varepsilon\|_{\CX^\alpha}.
$$
\end{lemma}

\medskip

\begin{proof}
Given any $u\in\CH^\beta$, we have $u=\Gamma\Gamma^{-1}(u)=\Gamma(u-\PT_u^sX_1-\PT_{\nabla u}^sX_2)$. We get
\begin{align*}
\|u-\Gamma\Gamma_\varepsilon^{-1}(u)\|_{\CH^\beta}&=\big\|\Gamma\big(u-\PT_u^sX_1-\PT_{\nabla u}^sX_2\big)-\Gamma\big(u-\PT_u^sX_1^{(\varepsilon)}-\PT_{\nabla u}^sX_2^{(\varepsilon)}\big)\big\|_{\CH^\beta}\\
&=\Big\|\Gamma\Big(\PT_u^s\big(X_1^{(\varepsilon)}-X_1\big)+\PT_{\nabla u}^s\big(X_2^{(\varepsilon)}-X_2\big)\Big)\Big\|_{\CH^\beta}\\
&\lesssim_{\EA,s}\left\|\PT_u^s\big(X_1^{(\varepsilon)}-X_1\big)+\PT_{\nabla u}^s\big(X_2^{(\varepsilon)}-X_2\big)\right\|_{\CH^\beta}\\
&\lesssim_{\EA,s}\|\EA_\varepsilon-\EA\|_{\CX^{2\alpha}}\|u\|_{L^2}
\end{align*}
since $s<s_\beta(\EA)$ implies the continuity of $\Gamma:\CH^\beta\to\CH^\beta$ and $X_i^{(\varepsilon)}-X_i$ depends linearly on $\EA_\varepsilon-\EA$ for $i\in\{1,2\}$. The result on $\Gamma-\Gamma_\varepsilon$ follows from the bound on $\Gamma_\varepsilon$ uniform with respect to $\varepsilon$.
\end{proof}

\medskip

This allows to prove density of the domain.

\medskip

\begin{corollary}
The domain $\CD_\EA$ is dense in $\CH^\beta$ for every $\beta\in[0,2\alpha)$.
\end{corollary}

\medskip

\begin{proof}
Given $f\in\CH^2$, $\Gamma(g_\varepsilon)\in\CD_\EA$ where $g_\varepsilon=\Gamma_\varepsilon^{-1}f\in\CH^2$ thus we can conclude with the previous Lemma that
$$
\lim_{\varepsilon\to0}\|f-\Gamma(g_\varepsilon)\|_{\CH^\beta}=0.
$$
The density of $\CH^2$ in $\CH^\beta$ completes the proof.
\end{proof}

\bigskip

\subsection{First properties of $H$}

Since $H$ is formally given by
$$
H=-\Delta+2iA\cdot\nabla+A\cdot A,
$$
we are able to define $(H,\CD_\EA)$ as an unbounded operator in $L^2$ associated to the enhanced potential $\EA$.

\medskip

\begin{definition}
We define $H:\CD_\EA\subset L^2\to L^2$ as
$$
Hu:=-\Delta u^\sharp+R(u)
$$
where $u^\sharp=\Phi(u)$ 
and
$$
R(u):=\P_{2iA}\nabla u+\PI(\nabla u,2iA)+\P_{A^2}u+\PI(u,A^2).
$$
\end{definition}

\medskip

The definition of $H$ is independant of the parameter $s\in(0,1)$. It is a very usefull tool to get differents bounds on the operator with the different representations
$$
Hu=-\Delta u_s^\sharp+R(u)+\Psi^s(u)
$$
where $u_s^\sharp:=\Phi^s(u)$ and
$$
\Psi^s(u):=-\Delta\big(\PT_u-\PT_u^s\big)X_1-\Delta\big(\PT_{\nabla u}-\PT_{\nabla u}^s\big)X_2\in C^\infty(\IT^2).
$$
For example, we can compare the graph norm of $H$
$$
\|u\|_H^2:=\|u\|_{L^2}^2+\|Hu\|_{L^2}^2
$$
and the natural norms of the domain
$$
\|u\|_{\CD_\EA}^2:=\|u\|_{L^2}^2+\|\Phi^s(u)\|_{\CH^2}^2
$$
with the following Proposition provided $s$ is small. Let $\beta:=\frac{1}{2}(\frac{4}{3}+2\alpha)$ and $\delta>0$. For $s\in(0,1)$ such that $s<s_\beta(\EA)$, we introduce the constant
$$
m_\delta^2(\EA,s):=ks^{\frac{\alpha-2}{2}}\|\EA\|_{\CX^\alpha}+k\delta^{-\frac{\beta}{2-\beta}}\left(\frac{\|\EA\|_{\CX^\alpha}}{1-m\frac{s^{\frac{2\alpha-\beta}{4}}}{\beta^*}\|\EA\|_{\CX^\alpha}}\right)^{\frac{2}{2-\beta}}\big(1+s^{\frac{\alpha}{2}}\|\EA\|_{\CX^\alpha}\big)
$$
with $k>0$ a large enough constant depending. In particular, $m_\delta^2(\EA,s)$ diverges as $s$ goes to $0$ or $s_\beta(\EA)$ or as $\delta$ goes to $0$.

\medskip

\begin{proposition}\label{H2estimate}
Let $u\in\CD_\EA$ and $s\in(0,1)$ such that $s<s_\beta(\EA)$. Then for any $\delta>0$, we have
$$
(1-\delta)\|u_s^\sharp\|_{\CH^2}\le\|Hu\|_{L^2}+m_\delta^2(\EA,s)\|u\|_{L^2}
$$
and
$$
\|Hu\|_{L^2}\le(1+\delta)\|u_s^\sharp\|_{\CH^2}+m_\delta^2(\EA,s)\|u\|_{L^2}
$$
with $u_s^\sharp=\Phi^s(u)$.
\end{proposition}

\medskip

\begin{proof}
Recall that for any $s\in(0,1)$, the operator is given by
$$
Hu=-\Delta u_s^\sharp+R(u)+\Psi^s(u)
$$
thus we need to bound $R$ and $\Psi^s$. For $u\in\CD_\EA$, we have
\begin{align*}
\|\P_{2iA}\nabla u+\PI(\nabla u,2iA)\|_{L^2}&\lesssim\|2iA\|_{\CC^{\alpha-1}}\|u\|_{\CH^\beta}\\
\|\P_{A^2}u+\PI(u,A^2)\|_{L^2}&\lesssim\|A^2\|_{\CC^{2\alpha-2}}\|u\|_{\CH^\beta}
\end{align*}
hence
$$
\|R(u)\|_{L^2}\lesssim\|\EA\|_{\CX^\alpha}\|u\|_{\CH^\beta}.
$$
We also have
$$
\|\Psi^s(u)\|_{L^2}\lesssim\|(\PT_u-\PT_u^s)X_1+(\PT_{\nabla u}-\PT_{\nabla u}^s)X_2\|_{\CH^2}\lesssim s^{\frac{\alpha-2}{2}}\|\EA\|_{\CX^\alpha}\|u\|_{L^2}.
$$
For $s<s_\beta(\EA)$, Proposition \ref{GammaBound} gives
$$
\|u\|_{\CH^\beta}\le\frac{1}{1-m\frac{s^{\frac{2\alpha-\beta}{4}}}{\beta^*}\|\EA\|_{\CX^\alpha}}\|u_s^\sharp\|_{\CH^\beta}
$$
thus we get
$$
\|Hu+\Delta u_s^\sharp\|_{L^2}\lesssim\frac{\|\EA\|_{\CX^\alpha}}{1-m\frac{s^{\frac{2\alpha-\beta}{4}}}{\beta^*}\|\EA\|_{\CX^\alpha}}\|u_s^\sharp\|_{\CH^\beta}+s^{\frac{\alpha-2}{2}}\|\EA\|_{\CX^\alpha}\|u\|_{L^2}.
$$
Since $0<\beta<2$, we have for any $t>0$
\begin{align*}
\|u_s^\sharp\|_{\CH^\beta}&\lesssim\left\|\int_0^t (-t'\Delta)e^{t'\Delta}u_s^\sharp\frac{\drm t'}{t'}\right\|_{\CH^\beta}+\left\|e^{t\Delta}u_s^\sharp\right\|_{\CH^\beta}\\
&\lesssim t^{\frac{2-\beta}{2}}\|u_s^\sharp\|_{\CH^2}+t^{-\frac{\beta}{2}}\|u_s^\sharp\|_{L^2}\\
&\lesssim t^{\frac{2-\beta}{2}}\|u_s^\sharp\|_{\CH^2}+t^{-\frac{\beta}{2}}\Big(1+s^{\frac{2\alpha}{4}}\|\EA\|_{\CX^\alpha}\Big)\|u\|_{L^2}.
\end{align*}
For any $\delta>0$, take
$$
t=\left(\frac{\delta\big(1-m\frac{s^{\frac{2\alpha-\beta}{4}}}{\beta^*}\|\EA\|_{\CX^\alpha}\big)}{k\|A\|_{\CX^\alpha}}\right)^{\frac{2}{2-\beta}}
$$
with $k$ the constant from the previous inequality. This yields 
$$
\|Hu+\Delta u_s^\sharp\|_{L^2}\lesssim m_\delta^2(\EA,s)\|u\|_{L^2}+\delta\|u_s^\sharp\|_{\CH^2}.
$$
and completes the proof.
\end{proof}

\medskip

\begin{remark}
In comparison to the work \cite{Mouzard} on the Anderson Hamiltonian
$$
u\mapsto-\Delta u+u\xi
$$
where the space white noise can be interpreted as an electric potential, one needs $s$ small for these bounds to hold here. In fact, one could perform the same kind of expansion with
$$
\Delta X_2=\P_{\nabla X_1}2iA
$$
at the price of a nonlinear dependance of $X_2$ with respect to $\EA$ in order to bypass the smallness condition on $s$. This would change the different bounds one get for $\Phi^s$ and $\Gamma$ but still yield a self-adjoint operator that is the limit of the regularised $H_\varepsilon$. Theorem $\textup{XIII}.26$ from \cite{RS1} guarantees that the different choice of construction coincide provided that $H$ is self-adjoint.
\end{remark}

\medskip

In particular, this implies that $(H,\CD_\EA)$ is a closed operator in $L^2$.

\medskip

\begin{proposition}
The operator $H$ is closed on its domain $\CD_\EA$.
\end{proposition}

\medskip

\begin{proof}
Let $(u_n)_{n\ge0}\subset\CD_\EA$ be a sequence such that 
$$
u_n\rightarrow u\quad\text{in }L^2\quad\text{and}\quad Hu_n\rightarrow v\quad\text{in }L^2.
$$
Proposition \ref{H2estimate} gives that $\big(\Phi^s(u_n)\big)_{n\ge0}$ is a Cauchy sequence in $\CH^2$ hence converges to $u_s^\sharp\in\CH^2$ for $s<s_\beta(\EA)$. Since $\Phi^s:L^2\to L^2$ is continuous, we have $\Phi^s(u)=u_s^\sharp$ hence $u\in\CD_\EA$. Finally, we have
\begin{align*}
\|Hu-v\|_{L^2}&\le\|Hu-Hu_n\|_{L^2}+\|Hu_n-v\|_{L^2}\\
&\lesssim_\EA\|u_s^\sharp-\Phi^s(u_n)\|_{\CH^2}+\|u-u_n\|_{L^2}+\|Hu_n-v\|_{L^2}
\end{align*}
hence $Hu=v$ and $H$ is closed on $\CD_\EA$.
\end{proof}

\medskip

We conclud this Section by computing the Hölder regularity of the functions in the domain.

\medskip

\begin{corollary*}
We have
$$
\CD_\EA\subset\CC^{1-\kappa}
$$
for any $\kappa>0$.
\end{corollary*}

\medskip

\begin{proof}
The Besov embedding in two dimensions implies
$$
\CH^2\hookrightarrow\CB_{\infty,2}^1\hookrightarrow L^\infty
$$
and $\Phi^s:L^\infty\to L^\infty$ is also invertible for $s$ small enough hence
$$
\CD_\EA=(\Phi^s)^{-1}(\CH^2)\subset L^\infty.
$$
First for $u\in\CD_\EA$, we have
\begin{align*}
\|u\|_{\CC^\alpha}&\lesssim\|u\|_{L^\infty}\|X_1\|_{\CC^\alpha}+\|\nabla u\|_{\CC^{-1}}\|X_2\|_{\CC^{\alpha+1}}+\|u^\sharp\|_{\CC^\alpha}\\
&\lesssim_\EA\|u\|_{L^\infty}+\|u^\sharp\|_{\CH^2}.
\end{align*}
Finally, this gives
\begin{align*}
\|u\|_{\CC^{1-\kappa}}&\lesssim\|u\|_{L^\infty}\|X_1\|_{\CC^{1-\kappa}}+\|\nabla u\|_{\CC^{\alpha-1}}\|X_2\|_{\CC^{2-\alpha+\kappa}}+\|u^\sharp\|_{\CC^{1-\kappa}}\\
&\lesssim_\EA\|u\|_{L^\infty}+\|u\|_{\CC^\alpha}+\|u^\sharp\|_{\CH^2}
\end{align*}
and the proof is complete.
\end{proof}

\bigskip

\subsection{Form comparison between $H$ and $\Delta$}

We have proven in Theorem \ref{H2estimate} that $Hu$ can be seen as a small perturbation of $-\Delta u^\sharp$ in norm. Here, we prove a similar statement in the quadratic form setting. Let $\eta:=\frac{\alpha}{4}$ and $\delta>0$. For $s\in(0,1)$ such that $s<s_{1-\eta}(\EA)$, define
$$
m_\delta^1(\EA,s):=(1+s^{\frac{\alpha-2}{2}})\|\EA\|_{\CX^\alpha}+\delta^{-\frac{1-\eta}{\eta}}\left(\frac{\|\EA\|_{\CX^\alpha}}{(1-\eta^{-1}s^{\frac{2\alpha+\eta-1}{4}}\|\EA\|_{\CX^\alpha})^2}\right)^{\frac{1}{\eta}}(1+s^{\frac{\alpha}{2}}\|\EA\|_{\CX^\alpha})
$$
with $k>0$ a large enough constant. In particular, $m_\delta^1(\EA,s)$ diverges as $s$ goes to $0$ or $s_{1-\eta}(\EA)$ or as $\delta$ goes to $0$.

\medskip

\begin{proposition}\label{H1estimate}
Let $u\in\CD_\EA$ and $s\in(0,1)$ such that $s<s_{1-\frac{\alpha}{4}}(\EA)$. For any $\delta>0$, we have
$$
(1-\delta)\langle\nabla u_s^\sharp,\nabla u_s^\sharp\rangle\le\langle u,Hu\rangle+m_\delta^1(\EA,s)\|u\|_{L^2}^2
$$
and
$$
(1-\delta)\langle\nabla u_s^\sharp,\nabla u_s^\sharp\rangle\le\langle u,H_\varepsilon u\rangle+m_\delta^1(\EA,s)\|u\|_{L^2}^2
$$
where $u_s^\sharp=\Phi^s(u)$.
\end{proposition}

\medskip

\begin{proof}
For $u\in\CD_\EA$, recall that
$$
Hu=-\Delta u_s^\sharp+R(u)+\Psi^s(u)
$$
where $u_s^\sharp=\Phi^s(u)\in\CH^2$. We have
\begin{align*}
\big\langle u,-\Delta u_s^\sharp\big\rangle&=\big\langle\PT_u^sX_1,-\Delta u_s^\sharp\big\rangle+\big\langle\PT_{\nabla u}^sX_2,-\Delta u_s^\sharp\big\rangle+\big\langle u_s^\sharp,-\Delta u_s^\sharp\big\rangle\\
&=-\big\langle\P_u^s\Delta X_1,u_s^\sharp\big\rangle-\big\langle\P_{\nabla u}^s\Delta X_2,u_s^\sharp\big\rangle+\big\langle\nabla u_s^\sharp,\nabla u_s^\sharp\big\rangle
\end{align*}
thus
$$
\langle u,Hu\rangle=-\big\langle\P_u^s\Delta X_1,u_s^\sharp\big\rangle-\big\langle\P_{\nabla u}^s\Delta X_2,u_s^\sharp\big\rangle+\big\langle\nabla u_s^\sharp,\nabla u_s^\sharp\big\rangle+\big\langle u,R(u)\big\rangle+\big\langle u,\Psi^s(u)\big\rangle.
$$
For $\eta\le\frac{\alpha}{2}$, we have
\begin{align*}
\big|\big\langle\P_u^s\Delta X_1,u_s^\sharp\big\rangle\big|&\lesssim\|\P_u^s\Delta X_1\|_{\CH^{2\alpha-2}}\|u_s^\sharp\|_{\CH^{1-\eta}}\lesssim\|\EA\|_{\CX^\alpha}\|u\|_{L^2}\|u_s^\sharp\|_{\CH^{1-\eta}},\\
\big|\big\langle\P_{\nabla u}^s\Delta X_2,u_s^\sharp\big\rangle\big|&\lesssim\|\P_{\nabla u}^s\Delta X_2\|_{\CH^{\alpha-1-\eta}}\|u_s^\sharp\|_{\CH^{1-\eta}}\lesssim\|\EA\|_{\CX^\alpha}\|u\|_{\CH^{1-\eta}}\|u_s^\sharp\|_{\CH^{1-\eta}},\\
\big|\big\langle u,\P_{2iA}\nabla u\big\rangle\big|&\lesssim\|u\|_{\CH^{1-\eta}}\|\P_{2iA}\nabla u\|_{\CH^{\alpha-1-\eta}}\lesssim\|\EA\|_{\CX^\alpha}\|u\|_{\CH^{1-\eta}}^2,\\
\big|\big\langle u,\P_{A^2}u+\PI(u,A^2)\big\rangle\big|&\lesssim\|u\|_{L^2}\|\P_{A^2}u+\PI(u,A^2)\|_{L^2}\lesssim\|\EA\|_{\CX^\alpha}\|u\|_{L^2}\|u\|_{\CH^{1-\eta}},\\
\big|\big\langle u,\Psi^s(u)\big\rangle\big|&\lesssim\|u\|_{L^2}\|(\P_u-\P_u^s)LX_1+(\P_{\nabla u}-\P_{\nabla u}^s)LX_2\|_{L^2}\lesssim s^{\frac{\alpha-2}{2}}\|\EA\|_{\CX^\alpha}\|u\|_{L^2}^2.
\end{align*}
The only term that is not a priori controlled is $\big\langle u,\PI(2iA,\nabla u)\big\rangle$ since the resonant term is singular if we only suppose that $u\in\CH^1$; this is where the almost duality property comes into play. We have
$$
\big\langle u,\PI(2iA,\nabla u)\big\rangle=\big\langle\P_u2iA,\nabla u\big\rangle+\DO(u,2iA,\nabla u)
$$
with the corrector $\DO(u,2iA,\nabla u)$ controlled if $u\in\CH^{1-\eta}$ with $\eta<\frac{\alpha}{2}$. The paraproduct is not singular however one can not use better regularity than $L^2$ for $u$ thus we use an integration by part to get
$$
\big\langle\P_u2iA,\nabla u\big\rangle=-\big\langle\textup{div}(\P_u2iA),u\big\rangle=-\big\langle\P_u\textup{div}(2iA),u\big\rangle+\big\langle\DB(u,2iA),u\big\rangle
$$
with
$$
\DB\big(a,(b_1,b_2)\big):=\textup{div}\big(\P_a(b_1,b_2)\big)-\P_a\textup{div}(b_1,b_2),
$$
see Appendix \ref{AppendixPC} for continuity estimates on $\DB$. We have
$$
\big\langle u,\PI(2iA,\nabla u)\big\rangle\lesssim\big|\big\langle\DB(u,2iA),u\big\rangle\big|+|\DO(u,2iA,\nabla u)|\lesssim\|\EA\|_{\CX^\alpha}\|u\|_{\CH^{1-\eta}}^2
$$
since $\textup{div}(A)=0$. Since $s<s_{1-\eta}(\EA)$, we get
$$
\big|\langle u,Hu\rangle-\langle\nabla u_s^\sharp,\nabla u_s^\sharp\rangle\big|\lesssim(1+s^{\frac{\alpha-2}{2}})\|\EA\|_{\CX^\alpha}\|u\|_{L^2}^2+\frac{\|\EA\|_{\CX^\alpha}}{\big(1-\eta^{-1}s^{\frac{2\alpha+\eta-1}{4}}\|\EA\|_{\CX^\alpha}\big)^2}\|u_s^\sharp\|_{\CH^{1-\eta}}^2.
$$
To complete the proof, one only has to interpolate the $\CH^{1-\eta}$ norm of $u_s^\sharp$ between its $\CH^1$ norm and its $L^2$ norm which is controlled by the $L^2$ norm of $u$, as in the proof of Proposition \ref{H2estimate}. Since $0<1-\eta<1$, we have for any $t>0$
\begin{align*}
\|u_s^\sharp\|_{\CH^{1-\eta}}&\lesssim\left\|\int_0^t (-t'\Delta)e^{t'\Delta}u_s^\sharp\frac{\drm t'}{t'}\right\|_{\CH^{1-\eta}}+\left\|e^{t\Delta}u_s^\sharp\right\|_{\CH^{1-\eta}}\\
&\lesssim t^{\frac{\eta}{2}}\|u_s^\sharp\|_{\CH^1}+t^{-\frac{1-\eta}{2}}\|u_s^\sharp\|_{L^2}\\
&\lesssim t^{\frac{\eta}{2}}\|u_s^\sharp\|_{\CH^1}+t^{-\frac{1-\eta}{2}}\Big(1+s^{\frac{2\alpha}{4}}\|\EA\|_{\CX^\alpha}\Big)\|u\|_{L^2}.
\end{align*}
For any $\delta>0$, take
$$
t=\left(\frac{\delta\big(1-\eta^{-1}s^{\frac{2\alpha+\eta-1}{4}}\|\EA\|_{\CX^\alpha}\big)^2}{k\|\EA\|_{\CX^\alpha}}\right)^{\frac{2}{\eta}}
$$
with $k$ the constant from the previous inequality. This yields 
$$
\big|\langle u,Hu\rangle-\langle\nabla u_s^\sharp,\nabla u_s^\sharp\rangle\big|\lesssim m_\delta^2(\EA,s)\|u\|_{L^2}+\delta\|u_s^\sharp\|_{\CH^1}.
$$
and completes the proof.
\end{proof}

\bigskip

\section{Self-adjointness and spectrum}\label{SelfAdj}

In this section, we prove that $H$ is self-adjoint with pure point spectrum. It is symmetric since $H\Gamma$ is the limit in norm of the regularised $H_\varepsilon\Gamma_\varepsilon$ as proved in Section \ref{Symmetric}. Hence it is enough to prove that 
$$
H+k:\CD_\EA\to L^2
$$ 
is surjective for some $k\in\IR$, this is the content of Section \ref{Selfadjoint}. In Section \ref{Resolventlimit}, we prove that $H_\varepsilon$ converges to $H$ in the stronger resolvent sense. Finally, we give in Section \ref{ComparisonHL} bounds for the eigenvalues of $H$ using the different representation 
$$
H=-\Delta\Phi^s+R+\Psi^s
$$
parametrised by $s\in(0,1)$ from the eigenvalues of $-\Delta$. In particular, it implies a Weyl-type law for $H$.

\bigskip

\subsection{The operator is symmetric}\label{Symmetric}

To prove that $H$ is symmetric, we use the regularised operator $H_\varepsilon$. Recall that $(H_\varepsilon,\CH^2)$ is self-adjoint and that $\Phi_\varepsilon^s:\CH^2\to\CH^2$ is continuous. In some sense, the operator $H$ should be the limit of 
$$
H_\varepsilon:=-\Delta+2iA_\varepsilon\cdot\nabla+A_\varepsilon^2
$$ 
as $\varepsilon$ goes to $0$ with $\EA_\varepsilon:=(A_\varepsilon,A_\varepsilon^2)$ a smooth approximation of $\EA$ in $\CX^\alpha$. Since $\CD(H_\varepsilon)=\CH^2$, one can not compare directly the operators. However given any $u\in L^2$, we have
$$
u=\big(\Gamma\circ\Phi^s\big)(u)=\lim_{\varepsilon\to0}\big(\Gamma_\varepsilon\circ\Phi^s\big)(u).
$$
Thus for $u\in\CD_\EA$, the approximation $u_\varepsilon:=\big(\Gamma_\varepsilon\circ\Phi^s\big)(u)
$ belongs to $\CH^2$ and one can consider the difference
$$
\|Hu-H_\varepsilon u_\varepsilon\|_{L^2}=\|(H\Gamma-H_\varepsilon\Gamma_\varepsilon)u^\sharp\|_{L^2}
$$
with $u^\sharp:=\Phi^s(u)$. The following Proposition assures the convergence of $H_\varepsilon\Gamma_\varepsilon$ to $H\Gamma$ provided $s<s_\beta(\EA)$ where $\beta=\frac{1}{2}(\frac{4}{3}+2\alpha)$.

\medskip

\begin{proposition}\label{Happrox}
Let $u\in\CD_\EA$ and $s\in(0,1)$ such that $s<s_\beta(\EA)$. Then
$$
\|Hu-H_\varepsilon u_\varepsilon\|_{L^2}\lesssim_{\EA,s}\|u_s^\sharp\|_{\CH^2}\|\EA-\EA_\varepsilon\|_{\CX^\alpha}
$$
with $u_s^\sharp=\Phi^s(u)$ and $u_\varepsilon:=\Gamma_\varepsilon u_s^\sharp$. In particular, this implies that $H_\varepsilon\Gamma_\varepsilon$ converges to $H\Gamma$ in norm as $\varepsilon$ goes to $0$ as operators from $\CH^2$ to $L^2$.
\end{proposition}

\medskip

\begin{proof}
We have
$$
H_\varepsilon u_\varepsilon=-\Delta u_s^\sharp+R_\varepsilon(u_\varepsilon)+\Psi_\varepsilon^s(u_\varepsilon)
$$
where $R_\varepsilon$ and $\Psi_\varepsilon^s$ are defined as $R$ and $\Psi^s$ with $\EA_\varepsilon$ instead of $\EA$. Since $\frac{4}{3}<\beta<2\alpha$, we have
\begin{align*}
\|R(u)-&R_\varepsilon(u_\varepsilon)\|_{L^2}\le\|R(u-u_\varepsilon)\|_{L^2}+\|(R-R_\varepsilon)(u_\varepsilon)\|_{L^2}\\
&\lesssim_{s,\EA}\|u-u_\varepsilon\|_{\CH^\beta}+\|\EA-\EA_\varepsilon\|_{\CX^\alpha}\|u_\varepsilon\|_{\CH^\beta}
\end{align*}
and
$$
\|\Psi^s(u)-\Psi_\varepsilon^s(u_\varepsilon)\|_{L^2}\lesssim_{s,\EA}\|u-u_\varepsilon\|_{L^2}+\|\EA-\EA_\varepsilon\|_{\CX^\alpha}\|u\|_{L^2}
$$
and the proof is complete since $s<s_\beta(\EA)$ implies
$$
\|u\|_{\CH^\beta}\lesssim_{s,\EA}\|u_s^\sharp\|_{\CH^\beta}.
$$
\end{proof}

\medskip

The symmetry of $H$ immediately follows.

\medskip

\begin{corollary*}
The operator $H$ is symmetric.
\end{corollary*}

\medskip

\begin{proof}
Let $u,v\in\CD_\EA$ and consider $u^\sharp:=\Phi^s(u)$ and $v^\sharp:=\Phi^s(v)$ for $s<s_\beta(\EA)$. Since $H_\varepsilon$ is a symmetric operator, we have
$$
\langle Hu,v\rangle=\lim_{\varepsilon\to0}\langle H_\varepsilon\Gamma_\varepsilon u_s^\sharp,\Gamma_\varepsilon v_s^\sharp\rangle=\lim_{\varepsilon\to0}\langle\Gamma_\varepsilon u_s^\sharp,H_\varepsilon\Gamma_\varepsilon v_s^\sharp\rangle=\langle u,Hv\rangle
$$
using that $H_\varepsilon\Gamma_\varepsilon$ converges to $H\Gamma$ and $\Gamma_\varepsilon$ to $\Gamma$ in norm convergence.
\end{proof}

\bigskip

\subsection{The operator is self-adjoint}\label{Selfadjoint}

In this Section, we prove that $(H,\CD_\EA)$ is self-adjoint. Being closed and symmetric, it is enough to prove that 
$$
(H+k)u=v
$$ 
admits a solution for some $k\in\IR$, see Theorem $\textup{X}.1$ in \cite{RS2}. This is done using the Babuška-Lax-Milgram Theorem, see \cite{Babuska}, and Theorem \ref{H1estimate} which implies that $H$ is almost surely bounded below for any $\delta\in(0,1)$ and $s$ small enough.

\medskip

\begin{proposition}\label{Resolvent}
Let $\delta\in(0,1)$ and $s\in(0,1)$ such that $s<s_{1-\frac{\alpha}{4}}(\EA)$. For $k>m_\delta^1(\EA,s)$, the operators $H+k$ and $H_\varepsilon+k$ are invertibles as unbounded operator in $L^2$. Moreover the operators
\begin{align*}
\big(H+k\big)^{-1}&:L^2\to\CD_\EA\\
\big(H_\varepsilon+k\big)^{-1}&:L^2\to\CH^2
\end{align*}
are bounded.
\end{proposition}

\medskip

\begin{proof}
Since $s<s_{1-\frac{\alpha}{4}}(\EA)$ and $k>m_\delta^1(\EA,s)$, Proposition \ref{H1estimate} gives
$$
\big(k-m_\delta^1(\EA,s)\big)\|u\|_{L^2}^2<\big\langle(H+k)u,u\big\rangle
$$
for $u\in\CD_\EA$. Considering the norm
$$
\|u\|_{\CD_\EA}^2=\|u\|_{L^2}^2+\|u_s^\sharp\|_{\CH^2}^2
$$
on $\CD_\EA$, this yields a weakly coercive operator using Proposition \ref{H2estimate} in the sense that
$$
\|u\|_{\CD_\EA}\lesssim_\EA\|(H+k)u\|_{L^2}=\sup_{\|v\|_{L^2}=1}\big\langle(H+k)u,v\big\rangle
$$
for any $u\in\CD_\EA$. Moreover, the bilinear map
$$
\left.\begin{array}{cccc}
B:&\CD_\EA\times L^2&\to&\IR\\
&(u,v)&\mapsto&\big\langle(H+k)u,v\big\rangle	
\end{array}\right.
$$
is continuous since Proposition \ref{H2estimate} implies
$$
\quad|B(u,v)|\le\|(H+k)u\|_{L^2}\|v\|_{L^2}\lesssim_\EA\|u\|_{\CD_\EA}\|v\|_{L^2}
$$
for $u\in\CD_\EA$ and $v\in L^2$. The last condition we need is that for any $v\in L^2\backslash\{0\}$, we have
$$
\sup_{\|u\|_{\CD_\EA}=1}|B(u,v)|>0.
$$
Let assume that there exists $v\in L^2$ such that $B(u,v)=0$ for all $u\in\CD_\EA$. Then
$$
\forall u\in\CD_\EA,\quad\langle u,v\rangle_{\CD_\EA,\CD_\EA^*}=0.
$$
hence $v=0$ as an element of $\CD_\EA^*$. By density of $\CD_\EA$ in $L^2$, this implies $v=0$ in $L^2$ hence the property we want. By the Theorem of Babuška-Lax-Milgram, for any $f\in L^2$ there exists a unique $u\in\CD_\EA$ such that
$$
\forall v\in L^2,\quad B(u,v)=\langle f,v\rangle.
$$
Moreover, we have $\|u\|_{\CD_\EA}\lesssim_\EA\|f\|_{L^2}$ hence the result for $(H+k)^{-1}$. The same argument works for $H_\varepsilon+k$ since Peroposition \ref{H1estimate} also holds for $H_\varepsilon$ with bounds uniform in $\varepsilon$.
\end{proof}

\medskip

As explained before, this immediatly implies that $H$ is self-adjoint. Moreover, the resolvent is a compact operator from $L^2$ to itself since $\CD_\EA\subset\CH^\beta$ for any $\beta\in[0,2\alpha)$ hence it has pure point spectrum.

\medskip

\begin{corollary}\label{SpectralResult}
The operator $H$ is self-adjoint with discret spectrum $\big(\lambda_n(\EA)\big)_{n\ge1}$ which is a nondecreasing diverging sequence without accumulation points. Moreover, we have
$$
L^2=\underset{n\ge1}{\bigoplus}\ \textup{Ker}\big(H-\lambda_n(\EA)\big)
$$
with each kernel being of finite dimension. We finally have the min-max principle
$$
\lambda_n(\EA)=\inf_D\sup_{u\in D;\|u\|_{L^2}=1}\langle Hu,u\rangle
$$
where $D$ is any $n$-dimensional subspace of $\CD_\EA$ that can also be written as
$$
\lambda_n(\EA)=\sup_{v_1,\ldots,v_{n-1}\in L^2}\ \inf_{\underset{\|u\|_{L^2}=1}{u\in\textup{Vect}(v_1,\ldots,v_{n-1})^\bot}}\langle Hu,u\rangle.
$$
\end{corollary}

\bigskip


\subsection{$H$ is the resolvent-limit of $H_\varepsilon$}\label{Resolventlimit}

Since the intersection of the domains of $H$ and $H_\varepsilon$ is trivial, the natural convergence of $H_\varepsilon$ to $H$ is in the resolvent sense, this is the following Proposition. In particular, this result explains why our operator $H$ is natural since the regularised operator satisfies
$$
\big(i\partial_1+A_1^{(\varepsilon)}\big)^2+\big(i\partial_2+A_2^{(\varepsilon)}\big)^2=H+c_\varepsilon+o_{\varepsilon\to0}(1)
$$
in the norm resolvent sense.

\medskip

\begin{proposition}\label{ResolventConvergence}
Let $\delta>0$ and $s\in(0,1)$ such that $s<s_\beta(\EA)$. Then for any constant $k>m_\delta^1(\EA,s)$, we have
$$
\|(H+k)^{-1}-(H_\varepsilon+k)^{-1}\|_{L^2\to L^2}\lesssim_{\EA,s}\|\EA-\EA_\varepsilon\|_{\CX^\alpha}.
$$
\end{proposition}

\medskip

\begin{proof}
Let $v\in L^2$. Since $H+k:\CD_\EA\to L^2$ is invertible, there exists $u\in\CD_\EA$ such that 
$$
v=(H+k)u
$$ 
thus
$$
\|(H+k)^{-1}v-(H_\varepsilon+k)^{-1}v\|_{L^2}=\|u-(H_\varepsilon+k)^{-1}(H+k)u\|_{L^2}.
$$
We introduce $u_\varepsilon:=\Gamma_\varepsilon\Phi^s(u)$ which converges to $u$ in $L^2$ and we have
$$
\|u-(H_\varepsilon+k)^{-1}(H+k)u\|_{L^2}\le\|u-u_\varepsilon\|_{L^2}+\|u_\varepsilon-(H_\varepsilon+k)^{-1}(H+k)u\|_{L^2}.
$$
Since Lemma \ref{LemmaGammaConvergence} gives
$$
\|u-u_\varepsilon\|_{L^2}\lesssim_{\EA,s}\|\EA-\EA_\varepsilon\|_{\CX^\alpha},
$$
we only have to bound the second term. We have
\begin{align*}
\|u_\varepsilon-(H_\varepsilon+k)^{-1}(H+k)u\|_{L^2}&=\|(H_\varepsilon+k)^{-1}\big((H_\varepsilon+k)u_\varepsilon-(H+k)u\big)\|_{L^2}\\
&\lesssim\|(H_\varepsilon+k)u_\varepsilon-(H+k)u\|_{L^2}\\
&\lesssim\|H_\varepsilon u_\varepsilon-Hu\|_{L^2}+k\|u_\varepsilon-u\|_{L^2}
\end{align*}
using Proposition \ref{Happrox}. In the end, we have
$$
\|(H+k)^{-1}v-(H_\varepsilon+k)^{-1}v\|_{L^2}\lesssim\|u_s^\sharp\|_{\CH^2}\|\EA-\EA_\varepsilon\|_{\CX^\alpha}
$$
hence the result since $(H+k)^{-1}:L^2\to\CD_\EA$ is continuous.
\end{proof}

\bigskip

\subsection{Comparison between the spectrum of $H$ and $\Delta$}\label{ComparisonHL}

The following Proposition provides sharp bounds for the eigenvalues of $H$ from the eigenvalues of $\Delta$, we denote by $(\lambda_n)_{n\ge1}$ the non-decreasing positive sequence of the eigenvalues of $-\Delta$ since it corresponds to the case $\EA=0$. For $\delta\in(0,1)$ and $s\in(0,1)$, introduce the constant
$$
m_\delta^+(\EA,s):=(1+\delta)(1+ms^{\frac{\alpha}{2}}\|\EA\|_{\CX^\alpha}).
$$
For $s<s_0(\EA)$, we also introduce
$$
m_\delta^-(\EA,s):=\frac{1-\delta}{1-ms^{\frac{\alpha}{2}}\|\EA\|_{\CX^\alpha}}.
$$
Recall $\beta=\frac{1}{2}(\frac{4}{3}+2\alpha)$. Write $a,b\le c$ to mean that we have both $a\le c$ and $b\le c$.

\medskip

\begin{proposition}\label{EVestimates}
Let $\delta\in(0,1)$ and $s\in(0,1)$ such that $s<s_\beta(\EA)\wedge s_{1-\frac{\alpha}{4}}(\EA)$. Given any $n\in\IZ^+$, we have
$$
\lambda_n(\EA),\lambda_n(\EA_\varepsilon)\le m_\delta^+(\EA,s)\lambda_n+2+2ms^{\frac{\alpha}{2}}\|\EA\|_{\CX^\alpha}+m_\delta^2(\EA,s).
$$
If moreover $s<s_0(\EA)$, we have
$$
\lambda_n(\EA),\lambda_n(\EA_\varepsilon)\ge m_\delta^-(\EA,s)\lambda_n-m_\delta^1(\EA,s).
$$
\end{proposition}

\medskip

\begin{proof}
Let $u_1^\sharp,\ldots,u_n^\sharp\in\CH^2$ be an orthonormal family of eigenfunctions of $-\Delta$ associated to $\lambda_1,\ldots,\lambda_n$ and consider
$$
u_i:=\Gamma u_i^\sharp\in\CD_\EA
$$
for $1\le i\le n$. Since $\Gamma$ is invertible, the family $(u_1,\ldots,u_n)$ is free thus the min-max representation of $\lambda_n(\EA)$ yields
$$
\lambda_n(\EA)\le\sup_{\underset{\|u\|_{L^2}=1}{u\in\textup{Vect}(u_1,\ldots,u_n)}}\langle Hu,u\rangle.
$$
Given any normalised $u\in\textup{Vect}(u_1,\ldots,u_n)$, we have
$$
\langle Hu,u\rangle\le\|Hu\|_{L^2}\le(1+\delta)\|u_s^\sharp\|_{\CH^2}+m_\delta^2(\EA,s)
$$
for $u_s^\sharp=\Phi^s(u)$ using Proposition \ref{H2estimate} and $s<s_\beta(\EA)$. Moreover
$$
\|u_s^\sharp\|_{\CH^2}\le(1+\lambda_n)\|u_s^\sharp\|_{L^2}\le(1+\lambda_n)\Big(1+ms^{\frac{\alpha}{2}}\|\EA\|_{\CX^\alpha}\Big)
$$
hence the upper bound
$$
\lambda_n(\EA)\le m_\delta^+(\EA,s)\lambda_n+2+2ms^{\frac{\alpha}{2}}\|\EA\|_{\CX^\alpha}+m_\delta^2(\EA,s).
$$
For the lower bound, we use the min-max representation of $\lambda_n(\EA)$ under the form
$$
\lambda_n(\EA)=\sup_{v_1,\ldots,v_{n-1}\in L^2}\ \inf_{\underset{\|u\|_{L^2}=1}{u\in\textup{Vect}(v_1,\ldots,v_{n-1})^\bot}}\langle Hu,u\rangle.
$$
Introducing
$$
F:=\textup{Vect}(u_m;m\ge n),
$$ 
we have that $F^\bot$ is a subspace of $L^2$ of finite dimension $n-1$ thus there exists a orthogonal family $(v_1,\ldots,v_{n-1})$ such that $F^\bot=\textup{Vect}(v_1,\ldots,v_{n-1})$. Since $F$ is a closed subspace of $L^2$ as an intersection of hyperplans, we have $F=\textup{Vect}(v_1,\ldots,v_{n-1})^\bot$ hence
$$
\lambda_n(\EA)\ge\inf_{\underset{\|u\|_{L^2}=1}{u\in F}}\langle Hu,u\rangle.
$$
Let $u\in F$ with $\|u\|_{L^2}=1$. Using Proposition \ref{H1estimate}, we have
\begin{align*}
\langle Hu,u\rangle&\ge(1-\delta)\langle\nabla u_s^\sharp,\nabla u_s^\sharp\rangle-m_\delta^1(\EA,s)\\
&\ge(1-\delta)\langle u_s^\sharp,-\Delta u_s^\sharp\rangle-m_\delta^1(\EA,s)\\
&\ge(1-\delta)\lambda_n\|u_s^\sharp\|_{L^2}^2-m_\delta^1(\EA,s).
\end{align*}
Finally using Proposition \ref{GammaBound} for $s<s_{1-\frac{\alpha}{4}}(\EA)$, we get
$$
\langle Hu,u\rangle\ge\frac{1-\delta}{1-ms^{\frac{\alpha}{2}}\|\EA\|_{\CX^\alpha}}\lambda_n-m_\delta^1(\EA,s)
$$
and the proof is complete.
\end{proof}

\medskip

In particular, taking
$$
s=\left(\frac{\delta}{m\|\EA\|_{\CX^\alpha}}\right)^{\frac{2}{\alpha}}
$$
gives the simpler bounds
$$
\lambda_n-m_\delta^1(\EA)\le\lambda_n(\EA)\le(1+\delta)^2\lambda_n+m_\delta^2(\EA)
$$
for any $\delta$ small enough. This is sharp enough to get an almost sure Weyl-type law from the Weyl law for the Laplacian $\Delta$, we denote by $N(\lambda)$ the number of eigenvalues of $-\Delta$ lower than $\lambda\in\IR$.

\medskip

\begin{corollary}
We have
$$
\lim_{\lambda\to\infty}\lambda^{-1}|\{n\ge0;\lambda_n(\EA)\le\lambda\}|=\pi
$$
\end{corollary}

\medskip

\begin{proof}
The lower and upper bounds on the eigenvalues give
$$
N\left(\frac{\lambda-m_\delta^2(\EA)}{1+\delta}\right)\le\big|\{n\ge0;\lambda_n(\EA)\le\lambda\}\big|\le N\big(\lambda+m_\delta^1(\EA)\big)
$$
hence the proof is complete using the result for the Laplacian.
\end{proof}

\bigskip

\section{Stochastic renormalisation of the singular potential}\label{Renormalisation}

The enhanced potential
$$
\EA:=(A,A^2)\in\CX^\alpha=\CC^{\alpha-1}\times\CC^{2\alpha-2}
$$
would be defined with $A^2:=A\cdot A$ for regular enough potential, for example $A\in L^\infty$ is enough. For arbitrary distributions $A\in\CC^{\alpha-1}$, this does not make sense since the product of two distributions in Hölder spaces can be defined only if the sum of their regularity exponents is positive. The magnetic Laplacian with white noise $B=\xi$ as magnetic field corresponds to this framework since the associated magnetic potential
$$
A:=\nabla^\bot\varphi
$$
where $\varphi=\Delta^{-1}\xi$ belongs to $\CC^{\alpha-1}$ for any $\alpha<1$. In this case, one has to make sense of
$$
A\cdot A=(-\partial_2\varphi)^2+(\partial_1\varphi)^2
$$
which is expected to belong to $\CC^{2\alpha-2}$ since $\alpha-1<0$. A natural way of proceeding is to consider a regularisation of the noise $\xi_\varepsilon:=\xi*\rho_\varepsilon$. In this case, the associated magnetic potential $A_\varepsilon$ is smooth and the product $A_\varepsilon\cdot A_\varepsilon$ is well-defined. The singularity of the limit translates as the almost sure divergence of the product as $\varepsilon$ goes to $0$. Indeed for $x\in\IT^2$, one has
\begin{align*}
\IE\big[A_\varepsilon(x)\cdot A_\varepsilon(x)\big]&=\IE\big[(\partial_2\Delta^{-1}\xi_\varepsilon)^2(x)+(\partial_1\Delta^{-1}\xi_\varepsilon)^2(x)\big]\\
&=\big\|\partial_2G_\varepsilon(x,\cdot)\big\|_{L^2}^2+\big\|\partial_1G_\varepsilon(x,\cdot)\big\|_{L^2}^2
\end{align*}
with $G$ the kernel of $\Delta^{-1}$ and $G_\varepsilon(x,\cdot):=G(x,\cdot)*\rho_\varepsilon$. In particular, $G$ is smooth on $\IT^2\backslash\{0\}$ with the same singularity as the Green function of the Laplacian at $x=0$. Hence the mean of the regularised product diverges as
$$
c_\varepsilon:=\IE[A_\varepsilon(0)\cdot A_\varepsilon(0)]\underset{\varepsilon\to0}{\sim}\frac{\ln(\varepsilon)}{4\pi^2}.
$$
While the mean of the random variable $A_\varepsilon(x)\cdot A_\varepsilon(x)$ diverges, one can try to describe the fluctuation around this asymptotic and find a limit to
$$
A_\varepsilon\cdot A_\varepsilon-\IE[A_\varepsilon\cdot A_\varepsilon]
$$
as $\varepsilon$ goes to $0$. It happens that this converges to a limit in the expected Hölder space $\CC^{2\alpha-2}$, this is the Wick product. As far as discrete associated models are concerned, this can be interpreted as a central limit Theorem where one describes the fluctuation around a diverging number of particules, see for example \cite{MP1} for an exemple with the Anderson Hamiltonian.

\medskip

\begin{theorem}
There exists a random distribution $A^2$ that belongs almost surely to $\CC^{2\alpha-2}$ such that
$$
\lim_{\varepsilon\to0}\IE\Big[\big\|A^2-(A_\varepsilon\cdot A_\varepsilon-c_\varepsilon)\big\|_{\CC^{2\alpha-2}}^p\Big]=0
$$
for any $p\ge1$.
\end{theorem}

\medskip

\begin{proof}
Let $(\Delta_k)_{k\ge-1}$ be the family of Paley-Littlewood projectors and consider
$$
a_\varepsilon:=A_i^{(\varepsilon)}
$$
for $i\in\{1,2\}$ and we want to show that $a_\varepsilon^2-\IE[a_\varepsilon^2]$ converges in the correct space as $\varepsilon$ goes to $0$. Gaussian hypercontractivity gives
$$
\IE\left[\big|\Delta_k(a_\varepsilon^2)\big|^p(x)\right]=\IE\left[\big|\Delta_k(a_\varepsilon^2)\big|^2(x)\right]^{\frac{p}{2}}
$$
hence it is enough to consider second order moment. Using the Wick formula, we have
\begin{align*}
\IE\left[\big|\Delta_k(a_\varepsilon^2)\big|^2(x)\right]&=\int_{\IT^2\times\IT^2}K_{\Delta_k}(x,y)K_{\Delta_k}(x,y)\IE\big[a_\varepsilon(y)^2a_\varepsilon(z)^2\big]\drm y\drm z\\
&=\int_{\IT^2\times\IT^2}K_{\Delta_k}(x,y)K_{\Delta_k}(x,y)\Big(\IE\big[a_\varepsilon^2(y)\big]\IE\big[a_\varepsilon^2(z)\big]+2\IE\big[a_\varepsilon(y)a_\varepsilon(z)\big]^2\Big)\drm y\drm z\\
&=\left(\int_{\IT^2}K_{\Delta_k}(x,y)\IE\big[a_\varepsilon^2(y)\big]\drm y\right)^2+2\int_{\IT^2\times\IT^2}K_{\Delta_k}(x,y)K_{\Delta_k}(x,y)\IE\big[a_\varepsilon(y)a_\varepsilon(z)\big]^2\drm y\drm z\\
&=\IE\Big[\big|\Delta_k(a_\varepsilon^2)\big|(x)\Big]^2+2I_\varepsilon(x).
\end{align*}
The first term corresponds to the diverging quantity that is substracted thus we only have to bound the second one. We have
\begin{align*}
I_\varepsilon(x)&\lesssim\int_{\IT^2\times\IT^2}K_{\Delta_k}(x,y)K_{\Delta_k}(x,y)\IE\big[\partial_i\Delta^{-1}\xi_\varepsilon(y)\partial_i\Delta^{-1}\xi_\varepsilon(z)\big]^2\drm y\drm z\\
&\lesssim\int_{\IT^2\times\IT^2}K_{\Delta_k}(x,y)K_{\Delta_k}(x,y)\big\langle\partial_iG_\varepsilon(y-\cdot),\partial_iG_\varepsilon(z-\cdot)\big\rangle^2\drm y\drm z\\
&\lesssim 2^{-k\delta}
\end{align*}
for any $\delta>0$. This implies that $\big(A_\varepsilon\cdot A_\varepsilon-c_\varepsilon\big)$ is a bounded family in $\CC^{2\alpha-2}$ for any $\alpha<1$. The same type of computations shows that $A_\varepsilon\cdot A_\varepsilon-c_\varepsilon$ actually converges to a distribution in $\CC^{2\alpha-2}$ that we denote by $A^2$. See for example Lemma $3.1$ in \cite{HairerLabbe}.
\end{proof}

\medskip

Then $\EA:=(A,A^2)$ belongs to $\CX^\alpha$ and the convergence in probability
$$
\lim_{\varepsilon\to0}\|\EA-\EA_\varepsilon\|_{\CX^\alpha}=0
$$
with 
$$
\EA_\varepsilon:=(A_\varepsilon,A_\varepsilon\cdot A_\varepsilon-c_\varepsilon)\in\CX^\alpha.
$$

\begin{remark}
\begin{itemize}
	\item The effect of a change of gauge $\widetilde A=A+\drm f$ on $H$ can be seen at the level of the regularised operator. It would give
	$$
	\widetilde H_\varepsilon=(i\partial_1+\widetilde A_1)^2+(i\partial_2+\widetilde A_2)^2-\tilde c_\varepsilon
	$$
	with
	$$
	\widetilde c_\varepsilon=\IE[\widetilde A_\varepsilon\cdot\widetilde A_\varepsilon].
	$$
	Since the spectral properties of the magnetic Laplacian with smooth potential are not affected by the choice of gauge, it only remains to see the impact in the renormalisation procedure. It gives
	$$
	\widetilde c_\varepsilon=\IE[A_\varepsilon\cdot A_\varepsilon]+2\IE[\drm f\cdot A_\varepsilon]+\IE[\drm f\cdot\drm f]=c_\varepsilon+\Delta f
	$$
	in the case of a deterministic change of gauge $\drm f$. This would change the spectral properties of the limit however one could recover the same operator by incorporating the term $\Delta f$ in the renormalisation procedure since it does not cause any divergence. The arbitrary choice one has to make in the renormalisation allows to deal with different gauge choice and should be motivated by the applications.
	\item As for the Anderson Hamiltonian, this raises interesting questions as far as probability is concerned. For example, the eigenvalues are random variables and a direct application of the method used in \cite{Mouzard} would give tail estimates. One could also consider the martingale problem associated to rough differential equations (RDEs) in the case of a time-independant distributional drift, see \cite{PZ} and the references therein.
\end{itemize}
\end{remark}

\bigskip

\appendix

\section{Paracontrolled calculus}\label{AppendixPC}

In this Appendix, we recall the basics for paracontrolled calculus based on the heat semigroup, see \cite{Mouzard} for details. See also \cite{BB3} where high order paracontrolled calculus was first introduced in a parabolic spacetime setting. We only consider the case of the Laplacian 
$$
L:=-\Delta
$$ 
but one can take any nice enough second order differential operators $L$ in Hörmander form
$$
L=-\sum_{i=1}^dV_i^2.
$$
Given any integer $b$, consider the families of operators
$$
Q_t^{(b)}:=\frac{(tL)^be^{tL}}{(b-1)!}\quad\text{and}\quad-\partial_tP_t^{(b)}=Q_t^{(b)}
$$
with $P_0^{(b)}=\textup{Id}$. Then there exists a polynomial $p_b$ of degree $b-1$ such that $P_t^{(b)}=p_b(tL)e^{tL}$ and we have
$$
f=\lim_{t\to0}P_t^{(b)}f=\int_0^1Q_t^{(b)}f\frac{\drm t}{t}+P_1^{(b)}f.
$$
The operators $Q_t^{(b)}$ and $P_t^{(b)}$ play respectively the role of Paley-Littlewood projector and Fourier series. Indeed, working on the torus gives
$$
\widehat{Q_t^{(b)}}(\lambda)=\frac{\left(t|\lambda|^2\right)^b}{(b-1)!}e^{-|\lambda|^2t}\quad\text{and}\quad\widehat{P_t^{(b)}}(\lambda)=p_b\left(t|\lambda|^2\right)e^{-|\lambda|^2t}
$$
hence $Q_t^{(b)}$ localizes in frequency around the annulus $|\lambda|\sim t^{-\frac{1}{2}}$ and $P_t^{(b)}$ localizes in frequency on the ball $|\lambda|\lesssim t^{-\frac{1}{2}}$. Since the measure $dt/t$ gives unit mass to each interval $[2^{-(i+1)},2^{-i}]$, the operator $Q_t^{(b)}$ is a multiplier that is approximately localized at `frequencies' of size $t^{-\frac{1}{2}}$. However, this decomposition using a continuous parameter does not satisfy a perfect cancellation property but the identity
$$
Q_t^{(b)}Q_s^{(b)}=\left(\frac{ts}{(t+s)^2}\right)^bQ_{t+s}^{(2b)}
$$
for any $s,t\in(0,1)$ which is small for $s\ll t$ or $t\ll s$ hence the parameter $b$ encodes a ``degree'' of cancellation. We denote $\mathsf{StGC}^k$ the standard family of Gaussian operators with cancellation of order $k$, see Section $1.2$ from \cite{Mouzard} for a precise definition.

\medskip

\begin{definition}
Given any $p,q\in[1,\infty]$ and $\alpha\in(-2b,2b)$, we define the Besov space $\CB_{p,q}^\alpha$ as the set of distribution $f\in\CD'$ such that
$$
\|f\|_{\CB_{p,q}^\alpha}:=\left\|e^{-L}f\right\|_{L_x^p}+\sup_{\underset{|\alpha|<k\le 2b}{Q\in\mathsf{StGC}^k}}\left\|t^{-\frac{\alpha}{2}}\|Q_tf\|_{L_x^p}\right\|_{L^q(t^{-1}\drm t)}<\infty.
$$
\end{definition}

\medskip

The exponent $p=q=2$ coresponds to Sobolev spaces and $p=q=\infty$ to Besov-Hölder spaces. As for the discrete Paley-Littlewood decomposition, this can be used to construct paraproducts. Formally, a product is described by
\begin{align*}
fg&=\lim_{t\to 0}P_t^{(b)}\left(P_t^{(b)}f\cdot P_t^{(b)}g\right)\\
&=\int_0^1\left\{Q_t^{(b)}\big(P_t^{(b)}f\cdot P_t^{(b)}g\big)+P_t^{(b)}\big(Q_t^{(b)}f\cdot P_t^{(b)}g\big)+P_t^{(b)}\big(P_t^{(b)}f\cdot Q_t^{(b)}g\big)\right\}\frac{\drm t}{t}\\
&\quad\quad\quad+P_1^{(b)}\left(P_1^{(b)}f\cdot P_1^{(b)}g\right).
\end{align*}
The family $P^{(b)}$ does not encode any cancellation while $Q^{(b)}$ encodes cancellation of order $2b$. We can transfer some of the cancellation from $Q^{(b)}$ to the $P^{(b)}$ in each term using the Leibnitz rule
$$
V_i(fg)=V_i(f)g+fV_i(g).
$$
Multiple uses of this allows to decompose the product as
$$
f\cdot g=\P_fg+\P_gf+\PI(f,g)
$$
where $\P_fg$ is a linear combination of
$$
\int_0^1Q_t^1\left(P_tf\cdot Q_t^2g\right)\frac{\drm t}{t}
$$
and $\PI(f,g)$ of
$$
\int_0^1P_t\left(Q_t^1f\cdot Q_t^2g\right)\frac{\drm t}{t}
$$
with $Q^1,Q^2\in\mathsf{StGC}^{\frac{b}{2}}$ and $P\in\mathsf{StGC}^{[0,b]}$. It is then a direct computation to prove the following continuity estimates. We refer to Appendix B in \cite{Mouzard} for a proof. The parameter $b$ can be taken as large as needed in applications.

\medskip

\begin{proposition}
Let $\alpha,\beta\in(-2b,2b)$ be regularity exponents.
\begin{itemize}
	\item[$\centerdot$] If $\alpha>0$, then $(f,g)\mapsto\P_fg$ is continuous from $\CC^\alpha\times\CH^\beta$ to $\CH^\beta$ and from $\CH^\alpha\times\CC^\beta$ to $\CH^\beta$.
	\item[$\centerdot$] If $\alpha<0$, then $(f,g)\mapsto\P_fg$ is continuous from $\CC^\alpha\times\CH^\beta$ to $\CH^{\alpha+\beta}$ and from $\CH^\alpha\times\CC^\beta$ to $\CH^{\alpha+\beta}$.
	\item[$\centerdot$] If $\alpha+\beta>0$, then $(f,g)\mapsto\PI(f,g)$ is continuous from $\CH^\alpha\times\CC^\beta$ to $\CH^{\alpha+\beta}$.
\end{itemize}
\end{proposition}

\medskip

Given any differentiable operator $D$ constructed from the $V_i$'s, one can consider a new paraproduct $\PT$ intertwined with $\P$ via the relation
$$
D\circ\PT=\P\circ D
$$
It enjoys the same properties as $\P$ and appear naturally in the weak formulation of PDEs involving $D$. In this work, we consider $\PT$ intertwinned with $\P$ via $L$, it is the natural object to describe the domain of $H$. The truncated paraproduct $\PT^s$ is defined as a linear combination of
$$
\int_0^sQ_t^1\left(P_tf\cdot Q_t^2g\right)\frac{\drm t}{t}.
$$
It is then possible to get explicit bounds on its convergence to $0$ as $s$ goes to $0$ and the difference $\PT-\PT^s$ is smooth since it is localised in ``frequency'' in $[s,1]$, see again Appendix $B$ in \cite{Mouzard} for details. In particular, there exists a universal constant $m>0$ that only depends on this construction such that
$$
\|\PT_a^sb\|_{\CH^\gamma}\le m\frac{s^{\frac{\beta-\gamma}{4}}}{\beta^*}\|a\|_{\CH^\alpha}\|b\|_{\CH^\beta}.
$$
for $s\in(0,1),\beta\in[0,2\alpha)$ and $\gamma\in[0,\beta)$. The last ingredient used in this work is continuity estimates on the corrector
$$
\DF(a,b,c)=\big\langle a,\PI(b,c)\big\rangle-\big\langle \P_ab,c\big\rangle
$$ 
and the commutator
$$
\DB\big(a,(b_1,b_2)\big)=\textup{div}\big(\P_a(b_1,b_2)\big)-\P_a\textup{div}(b_1,b_2).
$$

\medskip

\begin{proposition}
\begin{itemize}
	\item[$\centerdot$] Let $\alpha,\beta,\gamma\in(-2b,2b)$ such that $\beta+\gamma<1$ and $\alpha+\beta+\gamma\ge0$. If $\alpha<1$, then $(a,b,c)\mapsto\DF(a,b,c)$ extends in a unique trilinear operator from $\CH^\alpha\times\CC^\beta\times\CC^\gamma$ to $\IR$.
	\item[$\centerdot$] Let $\alpha,\beta\in(-2b,2b)$ such that. If $\alpha<1$, then $(a,b)\mapsto\DB(a,b)$ extends in a unique bilinear operator from $\CH^\alpha(\IT^2,\IR)\times\CC^\beta(\IR^2,\IR^2)$ to $\CH^{\alpha+\beta-1}(\IT^2,\IR)$.
\end{itemize}
\end{proposition}

\medskip

\begin{proof}
The proof of the continuity estimates of $\DF$ is already present in \cite{Mouzard} hence we only consider here $\DB$. We have
$$
\DB\big(a,(b_1,b_2)\big)=\partial_1\P_ab_1+\partial_2\P_ab_2-\P_a\partial_1b_1+\P_a\partial_2b_2
$$
hence the result follows from the continuity estimates on the commutators
$$
(a,b)\mapsto V_i\P_ab-\P_aV_ib,
$$
see for example Theorem $7$ in Section $2.2$ from \cite{BM} for the result on Hölder spaces. The extension to Sobolev spaces can be done as in \cite{Mouzard}.
\end{proof}

\bigskip

\noindent \textcolor{gray}{$\bullet$} L. Morin --  Univ. Rennes, CNRS, IRMAR - UMR 6625, F-35000 Rennes, France   \\
{\it E-mail}: leo.morin@univ-rennes1.fr

\noindent \textcolor{gray}{$\bullet$} A. Mouzard --  Univ. Rennes, CNRS, IRMAR - UMR 6625, F-35000 Rennes, France   \\
{\it E-mail}: antoine.mouzard@ens-rennes.fr

\end{document}